\documentclass[aps,prl,onecolumn,superscriptaddress]{revtex4-2} 

\usepackage{graphicx}  
\usepackage{dcolumn}   
\usepackage{bm}        
\usepackage{amssymb}   
\usepackage{siunitx}   
\usepackage{lipsum}
\usepackage{mathrsfs} 
\usepackage{hyperref}  

\begin{document}

\title{A new upper limit on the axion-photon coupling with an extended CAST run with a Xe-based Micromegas detector}

\author{K. Altenmüller}\affiliation{Centro de Astropartículas y Física de Altas Energías (CAPA) \& Departamento de Física Teórica, University de Zaragoza, 50009 — Zaragoza, Spain}
\author{V. Anastassopoulos}\affiliation{Physics Department, University of Patras, Patras, Greece}
\author{S. Arguedas-Cuendis}\affiliation{European Organization for Nuclear Research (CERN), 1211 Geneva 23, Switzerland}
\author{S. Aune}\affiliation{IRFU, CEA, Universit\'e Paris-Saclay, 91191 Gif-sur-Yvette, France}
\author{J. Baier}\affiliation{Physikalisches Institut, Albert-Ludwigs-Universität Freiburg, 79104 Freiburg, Germany}
\author{K. Barth}\affiliation{European Organization for Nuclear Research (CERN), 1211 Geneva 23, Switzerland}
\author{H. Bräuninger}\thanks{Deceased}\affiliation{Max-Planck-Institut für Extraterrestrische Physik, Garching, Germany}
\author{G. Cantatore}\affiliation{University of Trieste and Instituto Nazionale di Fisica Nucleare (INFN), Sezione di Trieste, Trieste, Italy}
\author{F. Caspers}\affiliation{European Organization for Nuclear Research (CERN), 1211 Geneva 23, Switzerland}\affiliation{European Scientific Institute, Archamps, France}
\author{J.F. Castel}\affiliation{Centro de Astropartículas y Física de Altas Energías (CAPA) \& Departamento de Física Teórica, University de Zaragoza, 50009 — Zaragoza, Spain}
\author{S.A. Çetin}\affiliation{Istinye University, Institute of Sciences, 34396, Sariyer, Istanbul, Turkey}
\author{F. Christensen}\affiliation{DTU Space, National Space Institute, Technical University of Denmark, 2800 Lyngby, Denmark}
\author{C. Cogollos}\affiliation{Institut de Ciències del Cosmos, Universitat de Barcelona (UB-IEEC), Barcelona, Catalonia, Spain}\affiliation{Centro de Astropartículas y Física de Altas Energías (CAPA) \& Departamento de Física Teórica, University de Zaragoza, 50009 — Zaragoza, Spain}
\author{T. Dafni}\affiliation{Centro de Astropartículas y Física de Altas Energías (CAPA) \& Departamento de Física Teórica, University de Zaragoza, 50009 — Zaragoza, Spain}
\author{M. Davenport}\affiliation{European Organization for Nuclear Research (CERN), 1211 Geneva 23, Switzerland}
\author{T.A. Decker}\affiliation{Lawrence Livermore National Laboratory, Livermore, California 94550, USA}
\author{K. Desch}\affiliation{Physikalisches Institut, University of Bonn, 53115 Bonn, Germany}
\author{D. Díez-Ibáñez}\affiliation{Centro de Astropartículas y Física de Altas Energías (CAPA) \& Departamento de Física Teórica, University de Zaragoza, 50009 — Zaragoza, Spain}
\author{B. Döbrich}\affiliation{European Organization for Nuclear Research (CERN), 1211 Geneva 23, Switzerland}
\author{E. Ferrer-Ribas}\affiliation{IRFU, CEA, Universit\'e Paris-Saclay, 91191 Gif-sur-Yvette, France}
\author{H. Fischer}\affiliation{Physikalisches Institut, Albert-Ludwigs-Universität Freiburg, 79104 Freiburg, Germany}
\author{W. Funk}\affiliation{European Organization for Nuclear Research (CERN), 1211 Geneva 23, Switzerland}
\author{J. Galán}\affiliation{Centro de Astropartículas y Física de Altas Energías (CAPA) \& Departamento de Física Teórica, University de Zaragoza, 50009 — Zaragoza, Spain}
\author{J.A. Garc\'ia}\affiliation{Centro de Astropartículas y Física de Altas Energías (CAPA) \& Departamento de Física Teórica, University de Zaragoza, 50009 — Zaragoza, Spain}
\author{A. Gardikiotis}\affiliation{Istituto Nazionale di Fisica Nucleare (INFN), Sezione di Padova, 35131 Padova, Italy}
\author{I. Giomataris}\affiliation{IRFU, CEA, Universit\'e Paris-Saclay, 91191 Gif-sur-Yvette, France}
\author{J. Golm}\affiliation{European Organization for Nuclear Research (CERN), 1211 Geneva 23, Switzerland}\affiliation{Institute for Optics and Quantum Electronics, Friedrich Schiller University Jena, Jena, Germany}
\author{C.H. Hailey}\affiliation{Physics Department and Columbia Astrophysics Laboratory, Columbia University, New York, New York 10027, USA}
\author{M.D. Hasinoff}\affiliation{Department of Physics and Astronomy, University of British Columbia, Vancouver,Canada}
\author{D.H.H. Hoffmann}\affiliation{Xi’An Jiaotong University, School of Science, Xi’An, 710049, China}
\author{I.G. Irastorza}\affiliation{Centro de Astropartículas y Física de Altas Energías (CAPA) \& Departamento de Física Teórica, University de Zaragoza, 50009 — Zaragoza, Spain}
\author{J. Jacoby}\affiliation{Physikalisches Institut, Albert-Ludwigs-Universität Freiburg, 79104 Freiburg, Germany}
\author{A.C. Jakobsen}\affiliation{DTU Space, National Space Institute, Technical University of Denmark, 2800 Lyngby, Denmark}
\author{K. Jakovčić}\affiliation{Rudjer Bošković Institute, Zagreb, Croatia}
\author{J. Kaminski}\affiliation{Physikalisches Institut, University of Bonn, 53115 Bonn, Germany}
\author{M. Karuza}\affiliation{Istituto Nazionale di Fisica Nucleare (INFN), Sezione di Trieste, Trieste, Italy}\affiliation{Faculty of Physics and Center for Micro and Nano Sciences and Technologies, University of Rijeka, 51000 Rijeka, Croatia}
\author{S. Kostoglou}\affiliation{European Organization for Nuclear Research (CERN), 1211 Geneva 23, Switzerland}
\author{C. Krieger}\affiliation{Universität Hamburg, Hamburg, Germany}
\author{B. Lakić }\thanks{Deceased}\affiliation{Rudjer Bošković Institute, Zagreb, Croatia}
\author{J.M .Laurent}\affiliation{European Organization for Nuclear Research (CERN), 1211 Geneva 23, Switzerland}
\author{G. Luzón}\affiliation{Centro de Astropartículas y Física de Altas Energías (CAPA) \& Departamento de Física Teórica, University de Zaragoza, 50009 — Zaragoza, Spain}
\author{C. Malbrunot}\affiliation{European Organization for Nuclear Research (CERN), 1211 Geneva 23, Switzerland}
\author{C. Margalejo}\email{Corresponding author: cmargalejo@unizar.es}\affiliation{Centro de Astropartículas y Física de Altas Energías (CAPA) \& Departamento de Física Teórica, University de Zaragoza, 50009 — Zaragoza, Spain}
\author{M. Maroudas}\affiliation{Institute of Experimental Physics, University of Hamburg, 22761 Hamburg, Germany}
\author{L. Miceli}\affiliation{Center for Axion and Precision Physics Research, Institute for Basic Science (IBS), Daejeon 34141, Republic of Korea}
\author{H. Mirallas}\affiliation{Centro de Astropartículas y Física de Altas Energías (CAPA) \& Departamento de Física Teórica, University de Zaragoza, 50009 — Zaragoza, Spain}
\author{P. Navarro}\affiliation{Department of Information and Communications Technologies, Technical University of Cartagena, 30203 — Murcia, Spain}
\author{L. Obis}\affiliation{Centro de Astropartículas y Física de Altas Energías (CAPA) \& Departamento de Física Teórica, University de Zaragoza, 50009 — Zaragoza, Spain}
\author{A. Özbey}\affiliation{Istinye University, Institute of Sciences, 34396, Sariyer, Istanbul, Turkey}\affiliation{Istanbul University-Cerrahpasa, Department of Mechanical Engineering, Avcilar, Istanbul, Turkey}
\author{K. Özbozduman}\affiliation{Istinye University, Institute of Sciences, 34396, Sariyer, Istanbul, Turkey}\affiliation{Boğaziçi University, Physics Department, Bebek, Istanbul, Turkey}
\author{T. Papaevangelou}\affiliation{IRFU, CEA, Universit\'e Paris-Saclay, 91191 Gif-sur-Yvette, France}
\author{O. Pérez}\affiliation{Centro de Astropartículas y Física de Altas Energías (CAPA) \& Departamento de Física Teórica, University de Zaragoza, 50009 — Zaragoza, Spain}
\author{M.J. Pivovaroff}\affiliation{Lawrence Livermore National Laboratory, Livermore, California 94550, USA}
\author{M. Rosu}\affiliation{Extreme Light Infrastructure - Nuclear Physics (ELI-NP), 077125 Magurele, Romania}
\author{E. Ruiz-Chóliz}\affiliation{Centro de Astropartículas y Física de Altas Energías (CAPA) \& Departamento de Física Teórica, University de Zaragoza, 50009 — Zaragoza, Spain}
\author{J. Ruz}\email{Corresponding author: Jaime.Ruz@cern.ch}\affiliation{Lawrence Livermore National Laboratory, Livermore, California 94550, USA}\affiliation{Centro de Astropartículas y Física de Altas Energías (CAPA) \& Departamento de Física Teórica, University de Zaragoza, 50009 — Zaragoza, Spain}
\author{S. Schmidt}\affiliation{Physikalisches Institut, University of Bonn, 53115 Bonn, Germany}
\author{M. Schumann}\affiliation{Physikalisches Institut, Albert-Ludwigs-Universität Freiburg, 79104 Freiburg, Germany}
\author{Y.K. Semertzidis}\affiliation{Center for Axion and Precision Physics Research, Institute for Basic Science (IBS), Daejeon 34141, Republic of Korea}\affiliation{Department of Physics, Korea Advanced Institute of Science and Technology (KAIST),
Daejeon 34141, Republic of Korea}
\author{S.K. Solanki}\affiliation{Max-Planck-Institut für Sonnensystemforschung, 37077 Göttingen, Germany}
\author{L. Stewart}\affiliation{European Organization for Nuclear Research (CERN), 1211 Geneva 23, Switzerland}
\author{T. Vafeiadis}\affiliation{European Organization for Nuclear Research (CERN), 1211 Geneva 23, Switzerland}
\author{J.K. Vogel}\affiliation{Lawrence Livermore National Laboratory, Livermore, California 94550, USA}\affiliation{Centro de Astropartículas y Física de Altas Energías (CAPA) \& Departamento de Física Teórica, University de Zaragoza, 50009 — Zaragoza, Spain}
\author{K. Zioutas}\affiliation{European Organization for Nuclear Research (CERN), 1211 Geneva 23, Switzerland}\affiliation{Physics Department, University of Patras, Patras, Greece}

\collaboration{CAST Collaboration}
\noaffiliation

\begin{abstract}
Hypothetical axions provide a compelling explanation for dark matter and could be emitted from the hot solar interior. The CERN Axion Solar Telescope (CAST) has been searching for solar axions via their back conversion to X-ray photons in a 9-T 10-m long magnet directed towards the Sun. We report on an extended run with the IAXO (International Axion Observatory) pathfinder detector, doubling the previous exposure time. The detector was operated with a xenon-based gas mixture for part of the new run, providing technical insights for future detector configurations in IAXO. No counts are detected in the 95\% signal-encircling region during the new run, while 0.75 are expected. The new data improve the axion-photon coupling limit to 5.8$\times 10^{-11}\,$GeV$^{-1}$ at 95\% C.L. (for $m_a \lesssim 0.02$\,eV), the most restrictive experimental limit to date.
\end{abstract}

\maketitle

\textit{Introduction.---}Very light pseudoscalar bosons, generically called axion-like particles (ALPs), appear in many motivated extensions of the Standard Model \cite{Ringwald:2012hr, Jaeckel:2010ni}. The paradigmatic example in this category is the axion, whose existence follows from the Peccei–Quinn mechanism as an explanation of why QCD (quantum chromodynamics) is perfectly time-reversal invariant within current experimental precision~\cite{Peccei:1977hh,Wilczek:1977pj,Weinberg:1977ma}. Axions and ALPs can be dark matter in the form of classical field oscillations that were excited in the early universe, by the re-alignment mechanism~\cite{Preskill:1982cy,Abbott:1982af,Dine:1982ah}, or by the decay of topological defects of the axion field~\cite{Marsh:2015xka}. 

There is a growing international program of experiments in search of these particles~\cite{Irastorza:2018dyq}. As dark matter components, they could be detected by a number of techniques, each of them optimized for a different axion mass $m_a$ range. Most notably, Sikivie-type axion haloscopes~\cite{Sikivie:1983ip} and in particular ADMX~\cite{ADMX:2021nhd}, have achieved sensitivity to QCD axion models in the range of $m_a\sim$~few~\si{\micro eV}. Independently of the dark matter assumption, axions can be produced and detected in the laboratory, as new forces mediated by them \cite{Arvanitaki:2014dfa} or
in light-shining-through-wall experiments like ALPSII at DESY~\cite{Oceano:2024maq}.
Axions can also be produced in stellar interiors, effectively draining energy and affecting the star's life span. These arguments provide restrictive limits on axion properties, and in some cases may even suggest new energy loss channels \cite{DiLuzio:2020wdo}. Axions produced in the Sun, offer another important opportunity for detection in the laboratory, in experiments dubbed axion helioscopes~\cite{Sikivie:1983ip}, the topic of this paper.

A most common strategy to search for axions relies on their generic two-photon coupling. It is given by the vertex
\begin{equation}
    {\mathscr{L}}_{a\gamma} =-\frac{1}{4}\,g_{a\gamma}
F^{\mu\nu}\widetilde F_{\mu\nu}a =g_{a\gamma} {\bf E}\cdot{\bf B}\,a
\end{equation}
where $a$ is the axion field, $F$ the electromagnetic field-strength tensor, $g_{a\gamma}$ the coupling constant, ${\bf E}$ the electric field and ${\bf B}$ the magnetic field.
This vertex enables the
decay $a\to\gamma\gamma$, as well as  the Primakoff production in stars, i.e., the
$\gamma\to a$ scattering in the Coulomb fields of charged particles in the
stellar plasma, and the coherent conversion $a\leftrightarrow\gamma$ in
laboratory or astrophysical $B$-fields \cite{Sikivie:1983ip,Raffelt:1987im}.

Solar axions can be produced in several processes, depending on their model-dependent interaction channels. We specifically consider axion production by Primakoff scattering of thermal photons deep in the Sun, a process that depends on the coupling constant $g_{a\gamma}$, which is also used for detection. Following this principle, axion helioscopes make use of a dipole magnet directed at the Sun to convert axions to X-rays.

This detection concept has been followed by the CERN Axion Solar Telescope (CAST), the most powerful axion helioscope so far built \cite{Anastassopoulos:2017ftl}. CAST was in operation at CERN from 2003 until 2021. During this time, the experiment has gone through different phases and has released a number of results, including 
a first
phase using evacuated magnet bores
\cite{Zioutas:2004hi, Andriamonje:2007ew}, followed by ``gas phases'' with $^4$He \cite{Arik:2008mq, Arik:2015cjv}
and $^3$He \cite{Aune:2011rx,Arik:2013nya}, to cover sensitivity to higher $m_a$ values.
Later on, CAST returned to evacuated magnet bores but with an improved detection line, dubbed IAXO (International Axion Observatory) pathfinder, that combined a new Micromegas detector with lower background levels, as well as a new X-ray telescope built specifically for axion searches~\cite{Aznar:2015iia} that is based on technology developed for the Nuclear Spectroscopic Telescope Array (NuSTAR) \cite{nustar2013}. This allowed CAST to produce what is at present the strongest experimental upper bound to the solar axion-photon coupling~\cite{Anastassopoulos:2017ftl}.

CAST also produced constraints to other (non-Primakoff) axion/ALP production channels in the Sun, \cite{Andriamonje:1186141, Barth:2013sma, CAST:2009klq}, as well as to chameleons \cite{Anastassopoulos2015} and hidden photons \cite{Redondo:2015iea}. In a later stage, CAST expanded its scope to the search for dark matter axions \cite{CAST:2020rlf, CAST-CAPP:2022} and solar chameleons via pressure sensing \cite{CAST-KWISP:2019}. 

In this last phase, the IAXO pathfinder line kept operating from September 2019 to June 2021, in order to improve statistics of the 2017 result. As a novelty, part of this data was taken with a new gas recirculation and filtering system to use a Xe-based gas mixture for the Micromegas detectors. The goal was to get lower background for the science analysis by testing this gas filling the Micromegas detector volume in real experiment conditions, which will be valuable for future Xe-based detectors. The analysis of the data taken in these runs is the main result of the present article. These new results are combined with the latest GridPix data from October 2017 to December 2018. The GridPix detector \cite{Krieger:2013cfa} is optimized for lower energy signals (i.e. it is more sensitive to axion-electron coupling $g_{ae}$) due to a lower energy threshold than Micromegas detectors as GridPix can detect individual electrons, and its results will be detailed in a forthcoming publication currently in preparation. Here these data only improved the result marginally in the upper limit for $g_{a\gamma}$. 

\textit{Experimental setup.---}The CAST helioscope makes use of a decommissioned prototype LHC magnet \cite{Zioutas:1998cc} of length \SI{9.26}{m} and a magnetic field of up to 9 T. It has two 4.3 cm diameter cold bores and can have detectors installed on both ends (sunrise and sunset side). It can track the Sun during sunrise and sunset for a total of $\sim\SI{3}{hours}$ per day. A more detailed description can be found in \cite{Andriamonje:2007ew,Anastassopoulos:2017ftl}, and some relevant differences with the current setup, and how they affect the overall efficiency, are described in the Supplemental Material \cite{supp}.

For the present data taking campaign, a single ultra-low background microbulk Micromegas detector \cite{Giomataris:1995fq, Andriamonje:2010zz} was installed on the sunrise side of the experiment. This detector is made from electroformed copper and Kapton, has a 3 cm drift distance, an X-ray transparent \SI{4}{\micro m} aluminized mylar window acting as cathode and a high granularity $6\times 6$\,cm stripped readout as anode, with $120\times120$ strips or channels of $0.5$\,mm pitch. Active and passive shieldings were installed. The active shielding consisted of a plastic scintillator placed above the detector for cosmic muons detection via coincidence.
The passive shielding was a lead box surrounding the detector with 10 to 15 cm thick walls to protect the detector from environmental gammas. The detector was coupled to the X-ray telescope, which is optimized for solar axion searches, maximizing its throughput in the ~3\,keV energy range. The whole line, including optics and detector, is described in \cite{Aznar:2015iia}.

In CAST, argon mixtures were used historically as gas in the Micromegas chamber due to being well studied and posing little engineering challenges \cite{Barth:2013sma,Anastassopoulos:2017ftl,SebastianTesis}, and it was also used during the first period of the data taking discussed in this work. However, argon's X-ray fluorescence around 3\,keV results in an increased background level at the energy range where the maximum of solar axion Primakoff flux is expected. Consequently, there has been a shift towards exploring alternative gas mixtures, particularly those based on xenon. Unlike the argon mixtures, which were utilized in an open loop system, xenon mixtures were employed in a closed loop system. Special attention has been given to potential contamination of water vapor or oxygen from leaks or outgassing. Implementing this change required the development of a new and more sophisticated gas recirculation system, incorporating moisture and oxygen filters, a recirculation pump and a buffer volume.
The gas mixtures used during the last data taking campaign were Ar + 2.3\% isobutane at 1.4 bar and 48.85\%Xe + 48.85\%Ne + 2.3\% isobutane at 1.05 bar. Xenon-based gas mixtures have a higher detection efficiency, allowing for the use of lower pressures that reduce the pressure difference between the vacuum pipe and the Micromegas gas volume, which would allow the use of thinner windows with a higher X-ray transparency, increasing the overall efficiency. We demonstrate that their use is possible, paving the way for the future use of even thinner windows and reduced gas pressures.

\textit{Data taking.---}Calibration of the detector was performed at different energies in the CAST X-ray tube at CERN using both Ar- and Xe-based mixtures, providing information on its response to photons of different energies that was later used for event discrimination.
During the data taking periods at CAST, calibrations using the 5.9\,keV peak of a $^{55}$Fe source were taken on a daily basis. This enabled the calibration of each background and tracking run in energy and the evaluation of the detector performance, stability and energy threshold over time. 
Once the detector was placed in CAST, the correct alignment of the full beamline was verified by placing a $\sim3$\,keV X-ray generator on the opposite side of the magnet. 
In addition to these measurements, targets were securely positioned during geometric surveys, allowing for verification of the detector's position. Furthermore, the Sun can be filmed twice per year with an optical telescope and a camera attached to the magnet, ensuring the pointing accuracy of the setup.

The data can be classified in three datasets as listed in Table \ref{tab:datasets}, two of them using argon and one using xenon. The argon data are split into two separate datasets because they were taken a few months apart and using different electronics parameters. Each of the three datasets has data obtained under axion-sensitive conditions -- namely during tracking, i.e., when the magnet is powered and oriented towards the Sun -- as well as data taken under background conditions, i.e., magnet powered on but not pointing to the Sun. The total tracking exposure is 314.6\,h and about 20 times more statistics is available for background.
\begin{table*}
\centering
\scriptsize
\caption{\label{tab:datasets} Datasets of the presented data taking campaign. Ar: 97.7\% Ar + 2.3\% isobutane at 1.4 bar; Xe: 48.85\% Xe + 48.85\% Ne + 2.3\% isobutane at 1.05 bar. The software efficiency $\epsilon_s$ at 5.9\,keV is indicated for each dataset. }
\begin{ruledtabular}
\begin{tabular}{ccccccc}
Dataset&Background exposure&Background level (2,7)\,keV &Tracking exposure&Gas&$\epsilon_s$ at 5.9\,keV&Years\\
&(h)&($\times10^{-6} \, \textnormal{keV}^{-1} \, \textnormal{cm}^{-2} \, \textnormal{s}^{-1}$)&(h) & &\\
\hline
1 & 2476 & $1.7\pm0.1$ & 130 & Ar & ~80\% & 2019 - 2020\\
2 & 335 & $2.3\pm0.4$& 25.6 & Ar & ~80\% & 2020\\
3 & 3416 & $1.5\pm0.1$& 159 & Xe & ~90\% & 2020 - 2021\\
\hline
Total & 6227 &  & 314.6 & & & \\
\end{tabular}
\end{ruledtabular}
\end{table*}
Dataset 3 uses optimized parameters of the electronics based on insights gained from the analysis of datasets 1 and 2. This allowed us to reach higher software efficiency $\epsilon_s$ by saving the information of all the readout channels each time there was a trigger instead of saving only the information of the channels over the energy threshold.

The detector was stable during most of the data taking campaign and showed an energy resolution of 20 to 23\% at 5.9\, keV. For datasets 1 and 2 the gain variation was 3.3\%, with an energy threshold consistently below 0.4\,keV. 
Dataset 3 required more attention to maintain optimal gas gain and transparency due to the recirculation system and its effect on the gas quality. Consequently, voltage and flow parameters were dynamically adjusted as required. The range of gain variation was 2.5\% to 9\%. Since the gain changes are constantly monitored they can be easily corrected with every calibration and they do not affect the data quality and analysis. The energy threshold remained below 0.75\,keV for most (83\%) of the runs.
A more detailed overview of the data taking stability is provided in the supplemental material \cite{supp}.

\textit{Data analysis and results.---}
The data analysis and background rejection was performed using the REST-for-physics framework \cite{REST,REST-git}, a ROOT-based collaborative software developed for data analysis and Geant4-based \cite{AGOSTINELLI2003250} simulations of rare event searches experiments and gaseous detectors using TPCs. This analysis was performed in a fully blind manner to prevent bias in the results.

The data analysis chain turns the raw data into events with a given energy and a physical position on the readout plane, for which topological observables can be computed. This allows for powerful background rejection as X-rays (small, symmetric, point-like and single-track events) are easily identified. Furthermore, events that occur in coincidence with a cosmic event in the active shielding are also removed. See \cite{MMcuts} and also the Supplemental Material \cite{supp} for an overview of the Micromegas events analysis and definition of the X-ray selection algorithms.

In this work, a background rejection taking into account calibrations taken at 6 different energies (1.5, 2.1, 3.0, 4.5, 5.9 and 8.0 keV) in an X-ray tube with interchangeable targets to produce the different energies, has been implemented. This approach is used instead of relying solely on the peak obtained in the daily calibrations with the $^{55}$Fe source at CAST. This allowed us to consider the energy dependence of the observables' distributions, and adapt the X-ray cuts for each energy range accordingly. This approach ensured that the figure of merit $\epsilon_{sg}/\sqrt{b}$, where $\epsilon_{sg}$ is the efficiency of the cuts when applied to the signal data and $b$ is the background rate after cuts, is maximized in every range independently. As a result, we achieved the highest software efficiencies to date using Micromegas detectors, ranging from 70\% to 90\% depending on the energy of the event. This improvement is especially noticeable at energies away from the 5.9\,keV peak of $^{55}$Fe, as we could now use information at other energies based on the runs in X-ray tube that help us better define X-ray-like events of any energy.
There have also been improvements in the hardware efficiency. A 4 \si{\micro\meter} polypropylene differential window previously installed in \cite{Anastassopoulos:2017ftl}, which had a low X-ray transparency at low energies, was removed for this run thus significantly increasing the efficiency at low energies. The use of Xe-based gas mixtures also provides a higher detection efficiency in the energy range of interest in addition to avoiding the argon fluorescence peak at $\sim3$\,keV. Furthermore, the detector response has been taken into account which also increases slightly the efficiency. Overall, the efficiency has been improved by a factor of 2 with respect to previous works, in particular~\cite{Anastassopoulos:2017ftl}, concentrated at low energies. A more detailed description of the improved efficiencies is provided in the Supplemental Material~\cite{supp}.

\begin{figure}[!htbp]
\includegraphics[width=0.5\columnwidth]{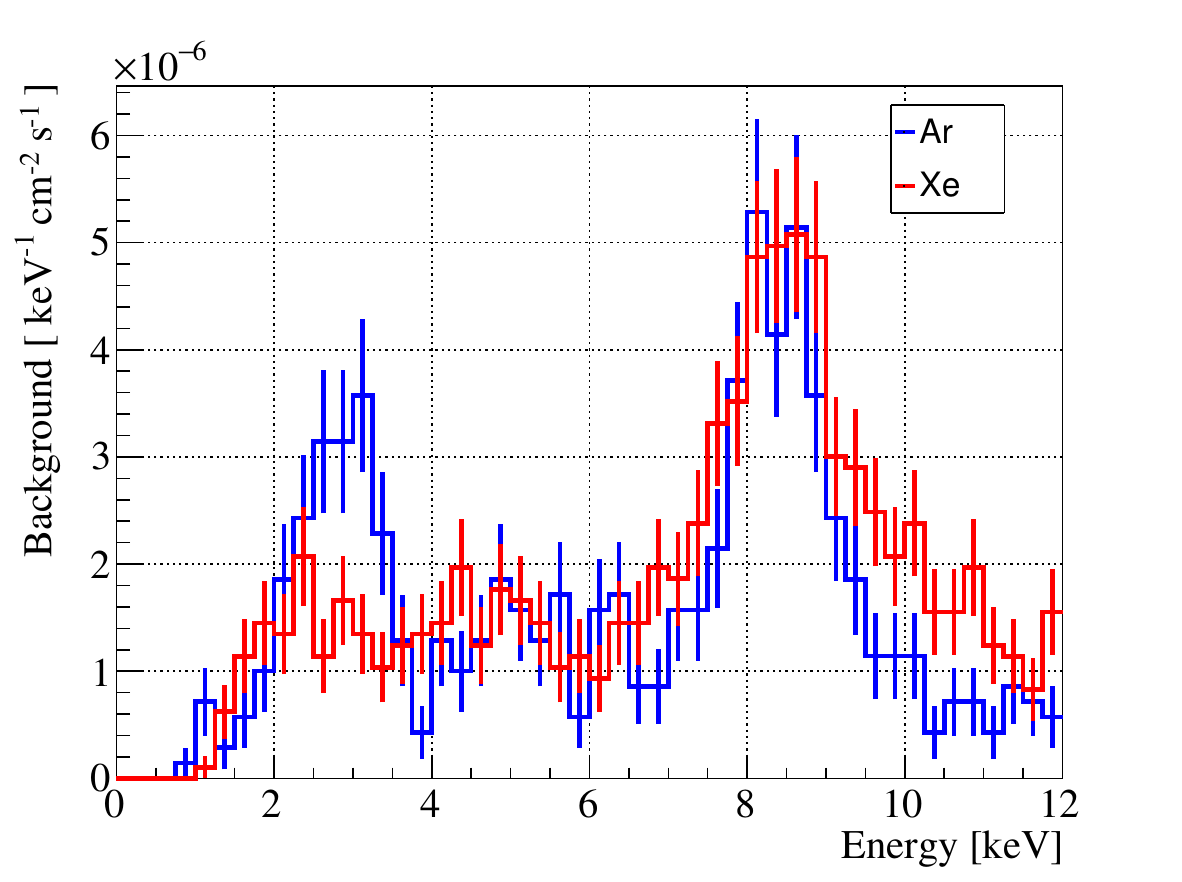}%
\caption{\label{fig:background}Energy distribution of background counts in the inner 20 mm diameter circular region of the detector in dataset 1 (blue line, labeled Ar) and dataset 3 (red line, labeled Xe). The $y$ axis is in normalized counts keV$^{-1}$~cm$^{-2}$~s$^{-1}$. The $~8$\,keV peak due to copper is seen in both spectra, but the $~3$\,keV Ar fluorescence peak is not present when using Xenon. This increases the signal-to-noise ratio in an energy region where the expected signal is maximal. }
\end{figure}

The background level achieved during the discussed campaign is based on 6227\,h of total data, divided in three different datasets listed in Table \ref{tab:datasets}. Most of the data belong to datasets 1 (Ar-based) and 3 (Xe-based), with a background level of $(1.7\pm 0.1)\times 10^{-6}$ keV$^{-1}$~cm$^{-2}$~s$^{-1}$ and $(1.5\pm 0.1)\times 10^{-6}$ keV$^{-1}$~cm$^{-2}$~s$^{-1}$ respectively. The background with xenon is the lowest in the energy region of interest due to the absence of the 3\,keV fluorescence peak (Fig. \ref{fig:background}), which is an important achievement for solar axion searches as it directly impacts the experiment's sensitivity, pushing it to the best levels to date. Dataset 2 is a smaller argon dataset in which we started saving the information of all the readout channels even if they were below the energy threshold. It has a slightly higher background rate of $(2.3\pm 0.4))\times 10^{-6}$ keV$^{-1}$~cm$^{-2}$~s$^{-1}$, but it is statistically compatible with the background level of dataset 1.

\begin{figure} [!b]
\includegraphics[width=0.5\columnwidth]{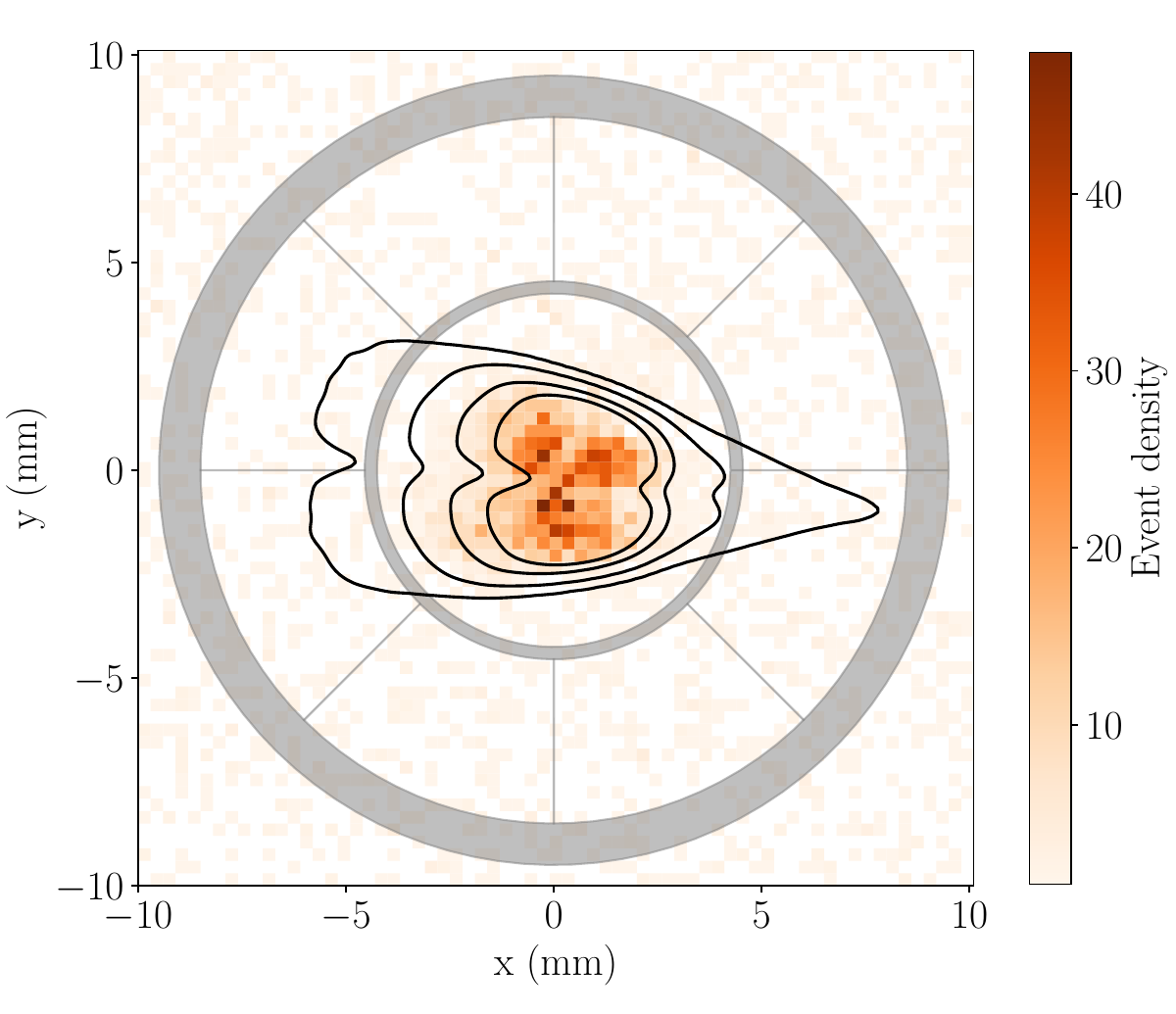}%
\caption{\label{fig:finger_alignment}2D hitmap of detected counts in the Micromegas detector plane during the calibration run with the source placed at the far end of the magnet, so the emitting X-rays cross the full beamline, and are focused by the optics. The distribution is overlaid with the 68\%, 85\%, 95\% and 99\% contours of the simulated ray-tracing (black lines). This data is used to determine the translation and rotation of the expected focused signal on the 2D detector plane.
The shade of the copper strongback of the X-ray window will block some of the signal and it is plotted as a grey shade.}
\end{figure}

\begin{figure}[!t]
\includegraphics[width=0.5\columnwidth]{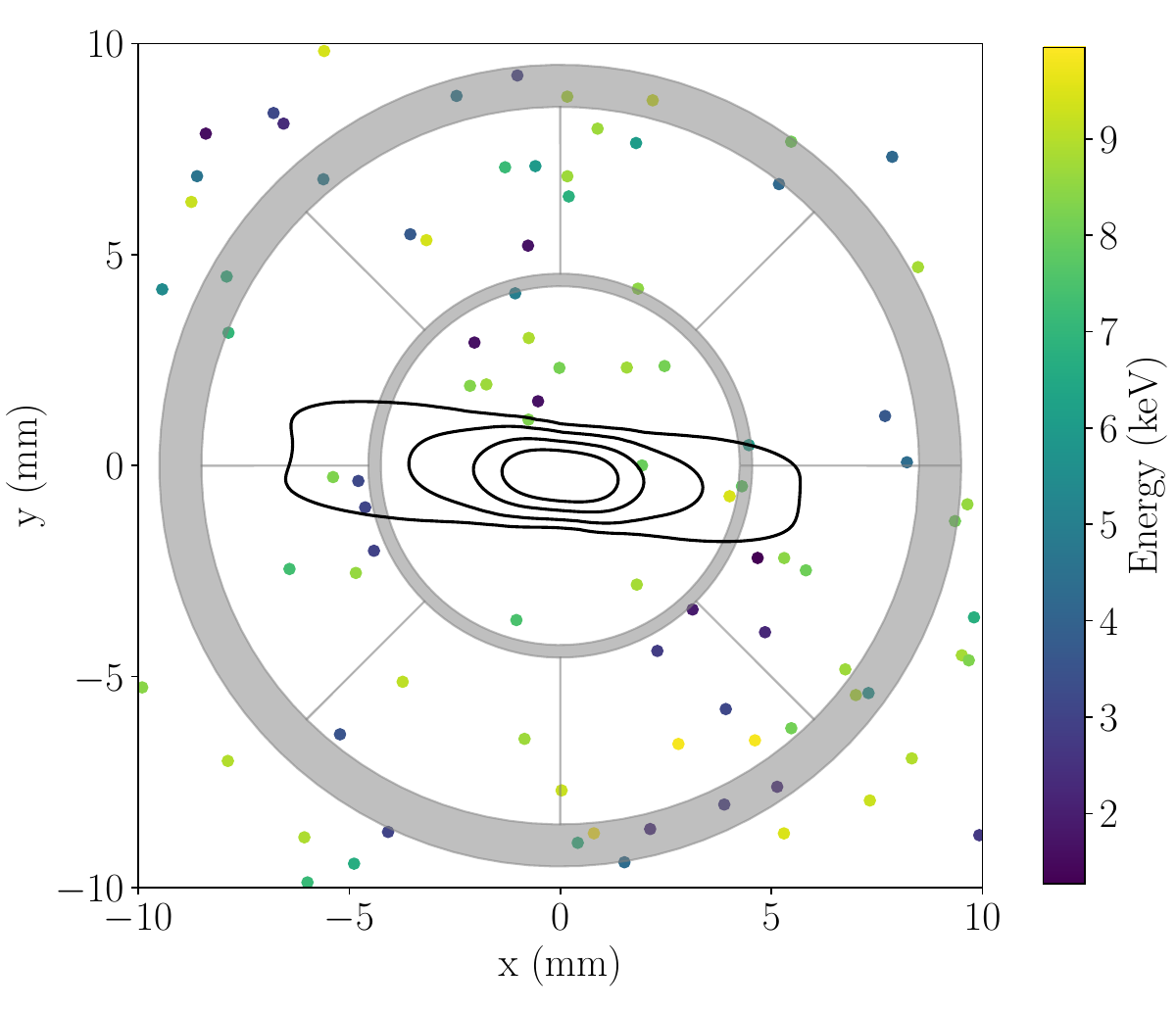}
\caption{\label{fig:axion_alignment}2D hitmap of the detected events during all the tracking runs in axion-sensitive conditions, during all three datasets considered in this paper. The colour of each dot represents its energy according to the scale on the right. The overlaid black lines represent the 68\%, 85\%, 95\% and 99\% signal encircling regions, according to the ray-tracing simulation of the optics. 
The window strongback is overlaid as a grey shade as in Fig.~\ref{fig:finger_alignment}}
\end{figure}

This background level is compared to the rate of X-ray-like observed events during the 314.6\,h of solar tracking time. The same selection algorithms are applied to these data and the aim is to look for any X-ray event excess during axion sensitive conditions. In order to do so, ray tracing simulations provide the information of the signal spot shape and position on the readout plane. The signal probability density function (PDF) is energy dependent, but most of the flux is in all cases focused into an area of a few mm$^2$.
The ray-tracing simulations were performed for a solar axion flux originating from an extended source with and angular size equivalent to the solar core, placed at infinity. Simulations were also conducted for an X-ray source like the one used for alignment calibrations, originating from a point source placed at ~10\,m distance. Comparing the latter with the real data allows to align the simulations density contours with the experimental data (Fig. \ref{fig:finger_alignment}). For this purpose, the contours containing 68\%, 85\%, 95\% and 99\% of the signal were computed.
The same alignment was applied to the X-ray events during tracking (Fig. \ref{fig:axion_alignment}), where we observe 0 counts in the energy range of interest (2 to 7 keV) in the 95\% signal encircling region in all 3 datasets, while expectations based on the background rate and exposure time is 0.75 counts.


\textit{Limit on the axion-photon coupling.---}
We use an unbinned likelihood method to compute an upper limit on $g_{a\gamma}$,
following the same methodology as in \cite{Aune:2011rx,Arik:2013nya,Arik:2015cjv,Anastassopoulos:2017ftl}, as it is better suited for low count experiments. The defined likelihood function is
\begin{equation}
    \ln \mathcal{L} = -R_T+\sum_{1}^{n} \ln R(E_i,\vec{x_i})
    \label{eq:like}
\end{equation}
where $R_T = s(g_{a\gamma}) + b$ is the sum of the expected number of signal counts $s(g_{a\gamma})$ based on the experimental setup, solar model and coupling constant, and background counts $b$ based on the background rate. The sum term goes over $n$ tracking events and it encodes the signal and background expected rates at energy $E_i$ and position $\vec{x_i}$ of event $i$, such that
\begin{equation}
    R(E_i,\vec{x_i}) = s(E_i,\vec{x_i}) + b(E_i).
\end{equation}
Here the background rate $b(E_i)$ is energy dependent (Fig. \ref{fig:background}) but it is considered spatially uniform, whereas the signal rate $s(E_i,\vec{x_i})$ depends on the position as well as the energy, as illustrated by the contours of the ray-tracing simulation in Fig. \ref{fig:axion_alignment}. This signal rate is given by 
\begin{equation}
    s(E_i,\vec{x_i}) = \frac{d\Phi_a}{dE}\cdot P_{a\rightarrow \gamma} \cdot \epsilon(E,\vec{x_i})
\end{equation}
where the detector response as a function of energy is encoded in the $\epsilon(E,\vec{x_i})$ term, which includes the X-ray optics efficiency of the  telescope, the hardware and software efficiencies, and the axion signal simulations defining the expected signal distribution (Fig. \ref{fig:axion_alignment}). 
The axion to photon conversion probability in an homogeneous magnetic field $B$ of length $L$ in vacuum is
\begin{equation}
P_{a \rightarrow \gamma} =\left(g_{a\gamma} B \frac{\sin\left(qL/2\right)}{q}\right)^2
\end{equation}
where $q=m_a^2/2E$ is the momentum transfer between axion and photon in vacuum. 
Finally, the differential Primakoff solar axion flux in keV$^{-1}$\,cm$^{-2}$\,s$^{-1}$ is given by the expression \cite{Andriamonje:2007ew}
\begin{equation}
    \frac{\mathrm{d}\Phi_a}{\mathrm{d}E}= 6.02\times 10^{10} g_{10}^2 \cdot E^{2.481} \cdot e^{-\frac{E}{1.205}}
\end{equation}
where $g_{10}=g_{a\gamma} / (10^{-10}\,\mathrm{GeV}^{-1})$ and energy $E$ is in keV.

The Bayesian posterior probability $\mathcal{P}(g_{a\gamma})$ is obtained from the likelihood function in Eq.~\ref{eq:like}, by $\mathcal{P} = \mathcal{L}\times\Pi$, where $\Pi(g_{a\gamma})$ is the prior probability which is chosen to be flat in $g_{a\gamma}^4$ for positive values, and $\Pi=0$ for negative ones. 

The resulting PDF is combined with results from \cite{Anastassopoulos:2017ftl}, which uses data up to 2015 and serves as the benchmark for the axion-photon coupling set by CAST. Additionally, data from the 2017-2018 campaign with a GridPix detector \cite{Krieger:2013cfa} \cite{SebastianTesis}, providing an extra 160 hours of data, is included. This combination allowed us to create an overall PDF that encapsulates all CAST data with sensitivity to $g_{a\gamma}$.

\begin{figure}[!htbp]
\includegraphics[width=0.5\columnwidth]{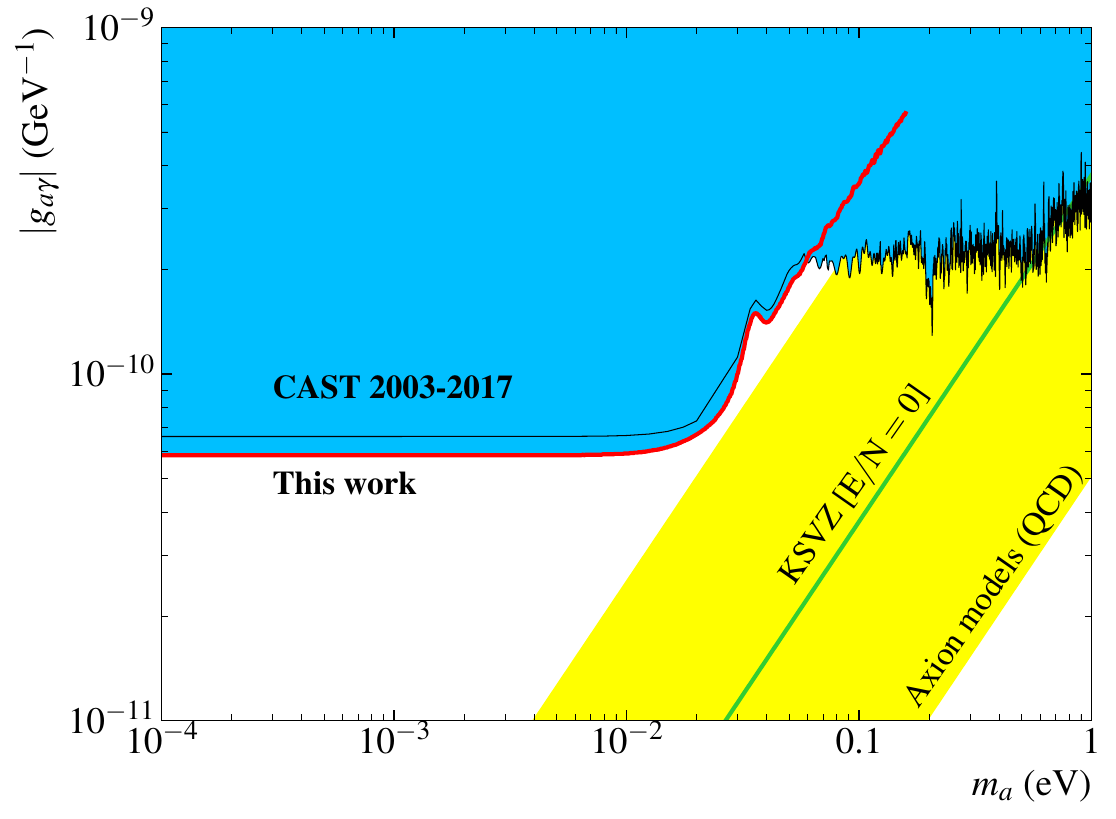}
\caption{\label{fig:helioscopesCAST} Parameter space for axions and ALPs showing the latest constraints in $g_{a\gamma}$. The red line indicates the region excluded by this work, which builds upon the former limit indicated in black. The yellow band represents the region QCD axion models point to.}
\end{figure}

Due to the absence of a significant excess of events over background, the data is consistent with no axion signal. Thus, an upper limit to $g_{a\gamma}$ is set by integrating the posterior probability $\mathcal{P}$ from 0 to 95\%. The analysis is repeated for different $m_a$ values to compute the exclusion line shown in red in Fig. \ref{fig:helioscopesCAST}, which is the resulting line of combining the current result with the aforementioned past data taking campaigns.

For $m_a \lesssim 0.02$\,eV, the upper limit is set to:
\begin{equation}
    g_{a\gamma} < 5.8 \times 10^{-11} \, \mathrm{GeV}^{-1} \; \mathrm{at\ } 95\% \; \mathrm{C.L.}
\end{equation}
For higher axion masses up to 0.06\,keV we have also set the most stringent limit, getting closer to the QCD axion band. For $m_a \gtrsim 0.06$\,eV, the data taken during the $^4$He and $^3$He as buffer gas periods is still the most competitive result from CAST \cite{Zioutas:2004hi, Andriamonje:2007ew,Arik:2008mq}.
The uncertainty in this latest result is dominated by the statistical effect of the low count rate available. The effect of systematic errors on this limit is discussed in the Supplemental Material \cite{supp}, which includes references \cite{JVogelTesis, 2101.08789}, but remains well below 10\% of the statistical uncertainty of the result. 

This result is a mild statistical underfluctuation as compared to the expected sensitivity of the experiment, defined as the median upper limit of multiple Monte Carlo simulations, which for the current exposure time is $g_{a\gamma}<0.59\times10^{-10}$\,GeV$^{-1}$. See the Supplemental Material \cite{supp}, which includes reference \cite{Cowan_2011}, for a more detailed explanation of the simulations of the experiment's potential sensitivity.

\textit{Conclusions.---}
The result presented in this paper represents the new best limit on $g_{a\gamma}$ in the range $m_a\lesssim0.02$~eV, superseding our previous best result, and going beyond the limit derived from the energy loss of horizontal branch (HB) stars ~\cite{Ayala:2014pea,Straniero:2015nvc}. Currently, our bound is surpassed only by astrophysical considerations on the  axion impact on the R2 parameter in globular clusters, which lead to $g_{a \gamma}<0.47 \times 10^{-10} \mathrm{GeV}^{-1}$. This last result, however, relies on the accurate counting of stars in globular clusters as well as on numerical simulations of stars in late evolutionary stages, and is marred by considerably larger uncertainties. 
This improvement originates not only from the additional statistics, but also from an improved detection efficiency, especially at low energies. This is in part thanks to the use of a Xe-based gas mixture in the Micromegas detector gas volume, something that also improves the background in the region of interest. This aspect constitutes a relevant technical achievement that provides useful operational experience for future similar implementations in BabyIAXO and IAXO. 

The improved energy threshold achieved and the more controlled low energy response of the detector are also of interest to search for other solar axion production channels at the lower energy range, like the $g_{ae}$-mediated ``ABC'' solar axions, something that will be the scope of forthcoming work. 

Given that CAST definitively stopped operation in 2021, the new bound on $g_{a\gamma}$ here presented will remain as the legacy result on solar axions, until new results come online once the BabyIAXO helioscope, now starting construction at DESY, starts producing data.

\textit{Acknowledgments.---}
We thank CERN for hosting the experiment and for technical support to operate the magnet and cryogenics. We thank V. Burwitz, G. Hartner of the MPE PANTER X-ray test facility for providing the opportunity to calibrate the X-ray telescope and for assistance in collecting and analysing the characterization data. We acknowledge support from NSERC (Canada); MSE (Croatia); CEA (France); BMBF (Germany) under the grant numbers 05 CC2EEA/9 and 05 CC1RD1/0, DFG (Germany) under grant numbers HO 1400/7-1 and EXC-153, GSRT (Greece), NSRF: Heracleitus II; Agence Nationale de la Recherche (France) ANR-19-CE31-133 0024; the Spanish Agencia Nacional de Investigación (AEI) under grants PID2019-108122GB-C31 and PID2022-137268NB-C51,funded by MCIN/AEI/10.13039/501100011033/FEDER, as well as the ``Planes complementarios, Programa de Astrof\'isica y F\'isica de Altas Energ\'ias'' from the ``European Union NextGenerationEU/PRTR'' funds; the European Research Council (ERC) under grants ERC-2017-AdG 788781 (IAXO+) and ERC 802836; the Marie Skłodowska-Curie grant agreement No 101026819 (LOBRES); Turkish Atomic Energy Authority (TAEK); NSF (USA) under Award No. 0239812; NASA under the grant number NAG5-10842; and IBS (Korea) with code IBS-R017-D1. Part of this work was performed under the auspices of the US Department of Energy by Lawrence Livermore National Laboratory under Contract No. DE-AC52-07NA27344 and funded through Lawrence Livermore National Laboratory Directed Research and Development project LDRD-17-ERD-030.

\clearpage

\begin{center}
\textbf{\large Supplemental Material}
\end{center}
\setcounter{equation}{0}
\setcounter{figure}{0}
\setcounter{table}{0}
\setcounter{page}{1}
\makeatletter
\renewcommand{\theequation}{S\arabic{equation}}
\renewcommand{\thefigure}{S\arabic{figure}}
\renewcommand{\thetable}{S\arabic{table}}

\section{Evolution and stability of the data}
\begin{figure}
    \includegraphics[width=0.5\columnwidth]    {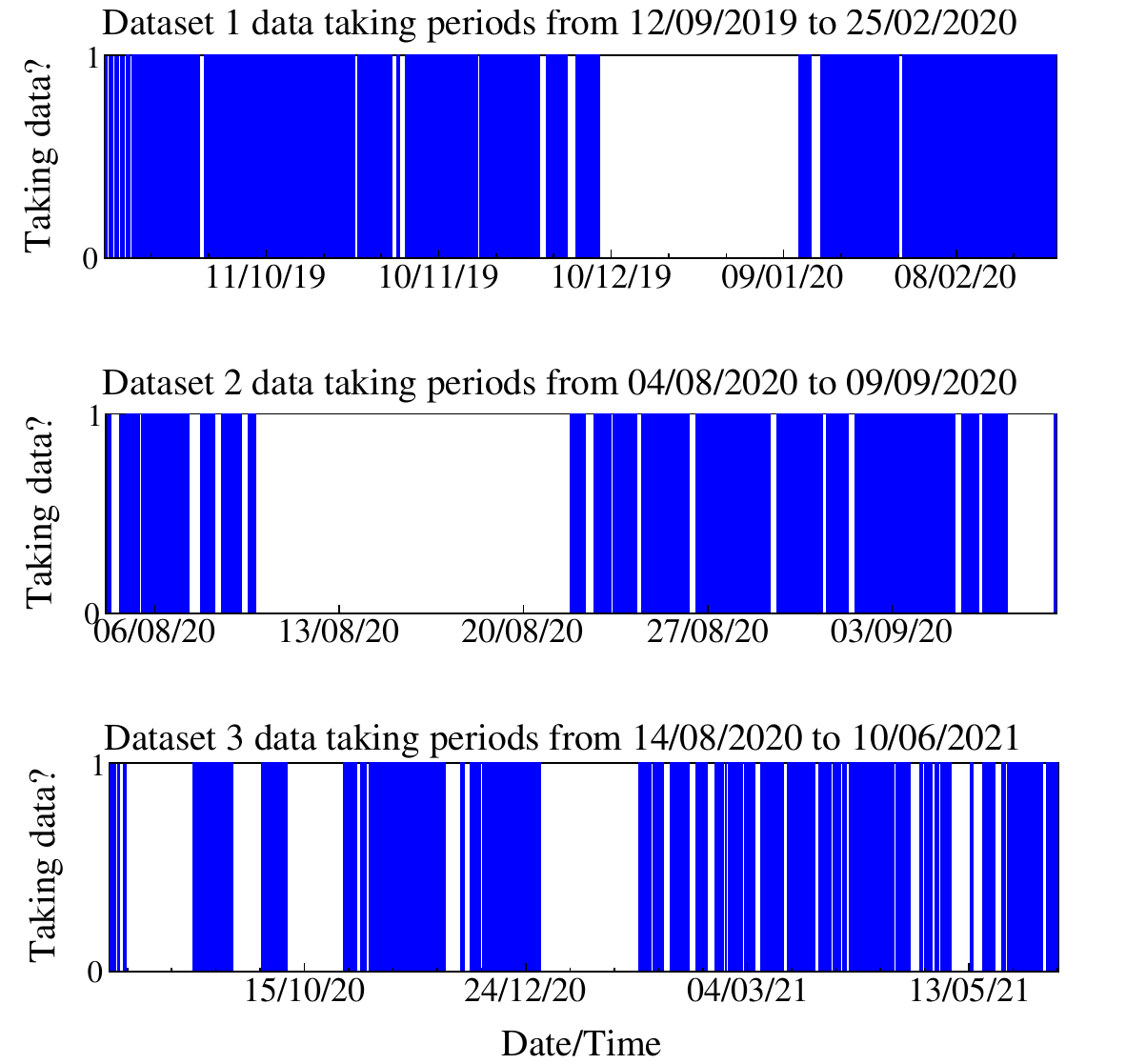}
    \caption{Data taking periods. Each subplot shows the full time extent of each dataset. The blue shaded regions are periods in which both tracking and background data were taken. During some of the white regions, background data was also taken.}
    \label{fig:periods}
\end{figure}

\begin{figure*}
\includegraphics[width=\textwidth]{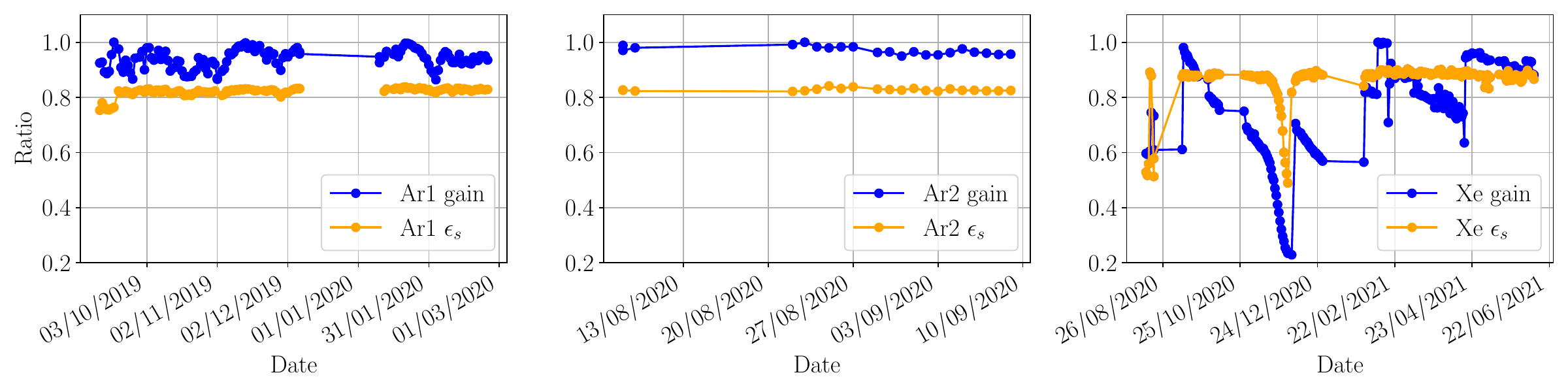}
\caption{\label{fig:ComnbinedStability}Time evolution of the gain based on the position of the $^{55}$Fe peak of the daily calibrations, expressed in relative terms, and of the software efficiency $\epsilon_s$. Left for dataset 1, centre for dataset 2, and right for dataset 3.}
\end{figure*}

\begin{figure}
\includegraphics[width=0.5\columnwidth]{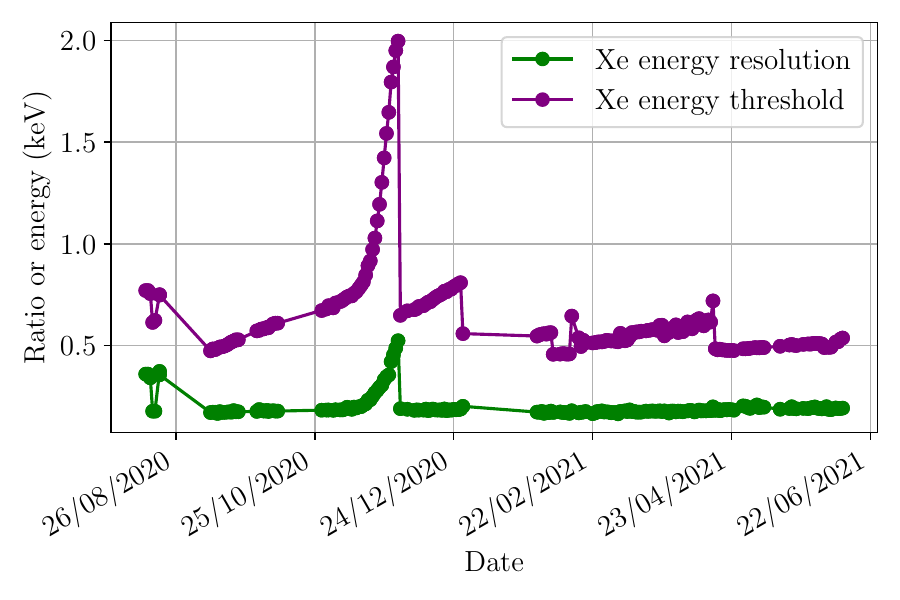}%
    \caption{Time evolution of the energy resolution and energy threshold in dataset 3. Each point corresponds to a calibration run, but not all of them have a tracking run associated. For example, of all the points with energy threshold above 1\,keV, only 6 do have data taken during tracking.}
    \label{fig:resAndTh}
\end{figure}

The added data in this analysis comprises a total of 314.6 hours of tracking and 6227 hours of background data. The data taking periods under axion sensitive conditions for each of the three datasets are illustrated in Fig.~\ref{fig:periods} as blue regions. The periods without solar tracking (white) were used to perform any necessary interventions and also to increase the background statistics.  The time evolution of some relevant parameters corresponding to each of the periods is depicted in Fig.\ref{fig:ComnbinedStability}.  During the two argon datasets (datasets 1 and 2), the gain was relatively unchanged with time and, most importantly, the software efficiency was very stable, meaning that despite the gain variations we were able to efficiently identify X-ray-like events in every run. The first few runs in dataset 1 have a slightly lower efficiency as the parameters of the electronics were still being optimized. It is worth noting that these efficiencies are calculated with respect to the number of events with a single track. However, 1-track events are ~95\% of the total events in the argon case, and ~97\% in the xenon case at 5.9 keV. 
For the xenon dataset there were strong gain variations. Our hypothesis is that these were due to the gas recirculation system affecting the gas quality in different ways:
\begin{itemize} 
    \item The moisture and oxygen filters get saturated with time, and have to be changed periodically.
    \item The moisture filter was found to be a radon emanator. This was introducing alpha particles in the gas that produced trips, so that the mesh voltage had to be modified accordingly and this directly effected the gain. No clear effect of the alphas in the low energy background was found.
    \item The total volume of the recirculation system was $\sim7$ liters. The mylar window permeation is different for each of the components of the gas (Xe, Ne and isobutane), leading to a small but steady change in the gas composition, which increases the relative amount of isobutane with time.
\end{itemize}
Each time the gain was low, some intervention was done: either changing filters, injecting fresh gas or modifying the voltages. Despite these gain variations, the efficiency of the X-ray cuts was close to 90\% during most of the runs, which means we were still able to identify X-ray-like events properly. However, during December 2020, the gain dropped to nearly 20\% of the initial gain. As shown in Fig. \ref{fig:resAndTh}, during this period the energy resolution at 5.9\,keV also increased to up to 50\%, and the energy threshold steadily went up to 2\,keV. Out of the 109 solar tracking runs taken with xenon mixtures, 6 have an energy threshold $>1$\,keV.

\section{Data analysis and event discrimination}

\begin{figure*}[tb]
\includegraphics[width=\textwidth]{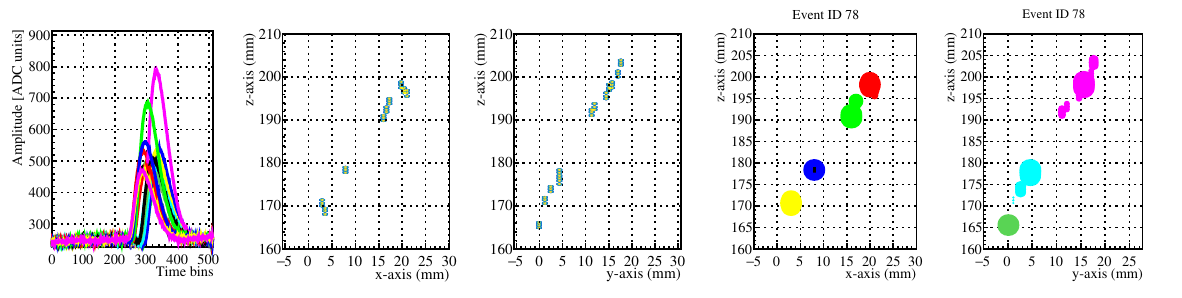}
\includegraphics[width=\textwidth]{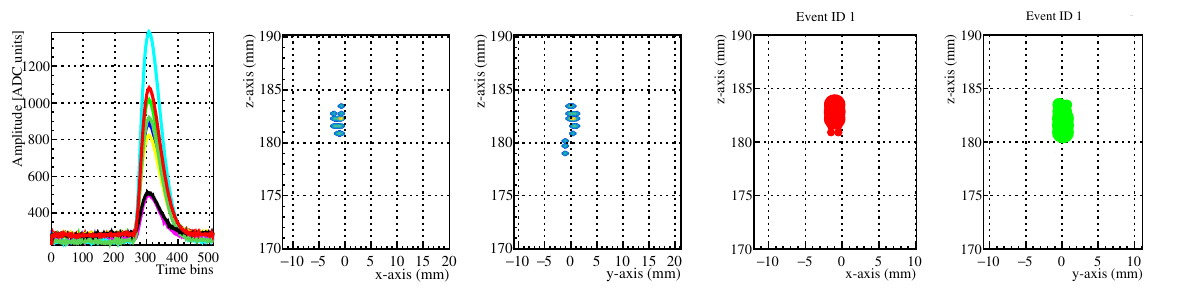}
\caption{\label{fig:Chain}Data analysis chain with the REST-for-physics framework, the top line for a background event and the bottom for a calibration event. From left to right, the \textit{signal event}, the \textit{hits event} in the XZ plane and in the YZ plane, and the \textit{track event} again in the XZ and YZ planes. Each colour in a \textit{track event} represents a different cluster.}
\end{figure*}

The data analysis has been performed with the REST-for-physics (Rare Event Searches Toolkit for Physics) framework \cite{RESTsupp}, and it can be fully replicated using the official version 2.4.0. The raw data in the form of voltage pulses are read from the detector and turned into a set of digitized waveforms, which are dubbed \textit{signal events} in REST. The shape of the raw signals depends on parameters such as the shaping time, trigger delay or time duration of the acquisition window. Enough bins without signal are left at the beginning of the window, to be able to compute a baseline. After some processing and taking into account the detector and readout descriptions, these events turn into \textit{hits events} with a given physical position and energy (and relative time). These hits can be grouped into \textit{track events} if they fulfill a set of conditions that classify them as belonging to the same physical event, such as the maximum 3-dimensional distance (XYZ, or equivalently XY-time). This analysis chain is depicted in Fig. \ref{fig:Chain}.

The use of a stripped readout allows for topological analysis of the \textit{track events}, which enables the definition of selection algorithms based on the topological shape of the events. As seen in Fig. \ref{fig:Chain}, X-ray like events are small, symmetric, point like and consist of a single track, whereas background events have multiple tracks and are highly asymmetric (e.g. muon or alpha particle tracks). The process of Micromegas data cut optimization used in this work is explained in \cite{MMcutssupp}. A further cut based on muons is applied. If an event in the Micromegas happens just after a muon signal is detected, the event is removed. Random coincidences are expected to be less than 0.5\%, giving this cut a very high efficiency while simultaneously reducing the background level by a factor $\sim1.7$ . 

\section{Efficiencies and detector response}
\begin{figure*}
\includegraphics[width=\textwidth]{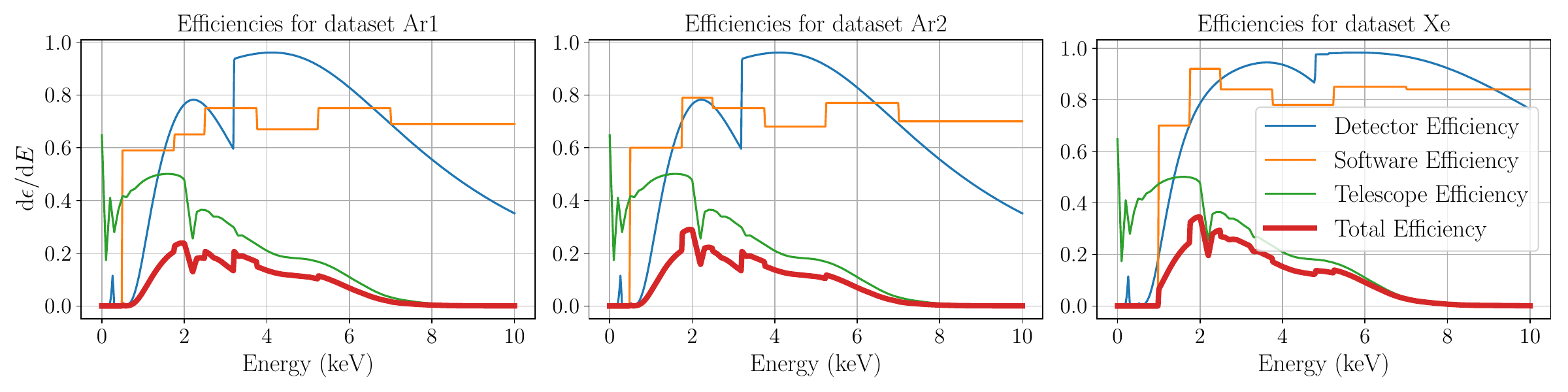}
\caption{\label{fig:Efficiencies}Depiction of the total efficiency for each dataset and the separate components: detector (meaning gas absorption and window transmission) in blue, software in orange and telescope in green. The energy threshold is taken into account as part of the software efficiency, and is represented as a sharp drop at 0.5 keV for Ar and 1 keV for Xe. The combined total efficiency is drawn in red. A horizontal line at 1 would represent 100\% efficiency.}
\end{figure*}

The efficiencies, illustrated in Fig. \ref{fig:Efficiencies}, are energy dependent and can be classified as follows:
\begin{itemize}
    \item Detector efficiency takes into account how many X-rays of each energy go through the mylar window and are absorbed in the gas. This efficiency is later corrected by the detector response as explained below.
    \item Software efficiency is defined for each energy range as the rate of calibration events that remain after applying the Micromegas and veto cuts. The energy threshold is included in this efficiency by setting it to 0 for energies below the energy threshold.
    \item Telescope efficiency is the efficiency of the optic measured at the MPE PANTER X-ray test facility in Munich in July 2016, and it is optimized for axion searches, maximizing the telescope throughput at low energies.
    \item Total efficiency is the product of all the efficiencies.
\end{itemize}

Monte Carlo simulations of the detector response have been performed with REST-for-physics \cite{RESTsupp} and Geant4 \cite{AGOSTINELLI2003250supp}.
A flat X-ray flux between 0 and 12 keV from a point source placed in the center of the copper pipe, 100\,mm away from the the detector window, was simulated.

The output spectrum from the simulation is a result of the photons that were transmitted through the 4 microns aluminized Mylar window and interacted with the gas in the chamber, thus depositing their energy fully or partially. If one only considers events that deposit all their energy to count towards the efficiency, there is an efficiency loss because if the energy of the incoming photon is high enough (e.g. $\gtrsim 3 $\,keV in Ar), it can produce a fluorescence peak and deposit less energy than the photon originally carried.

\begin{figure}
\includegraphics[width=0.5\columnwidth]{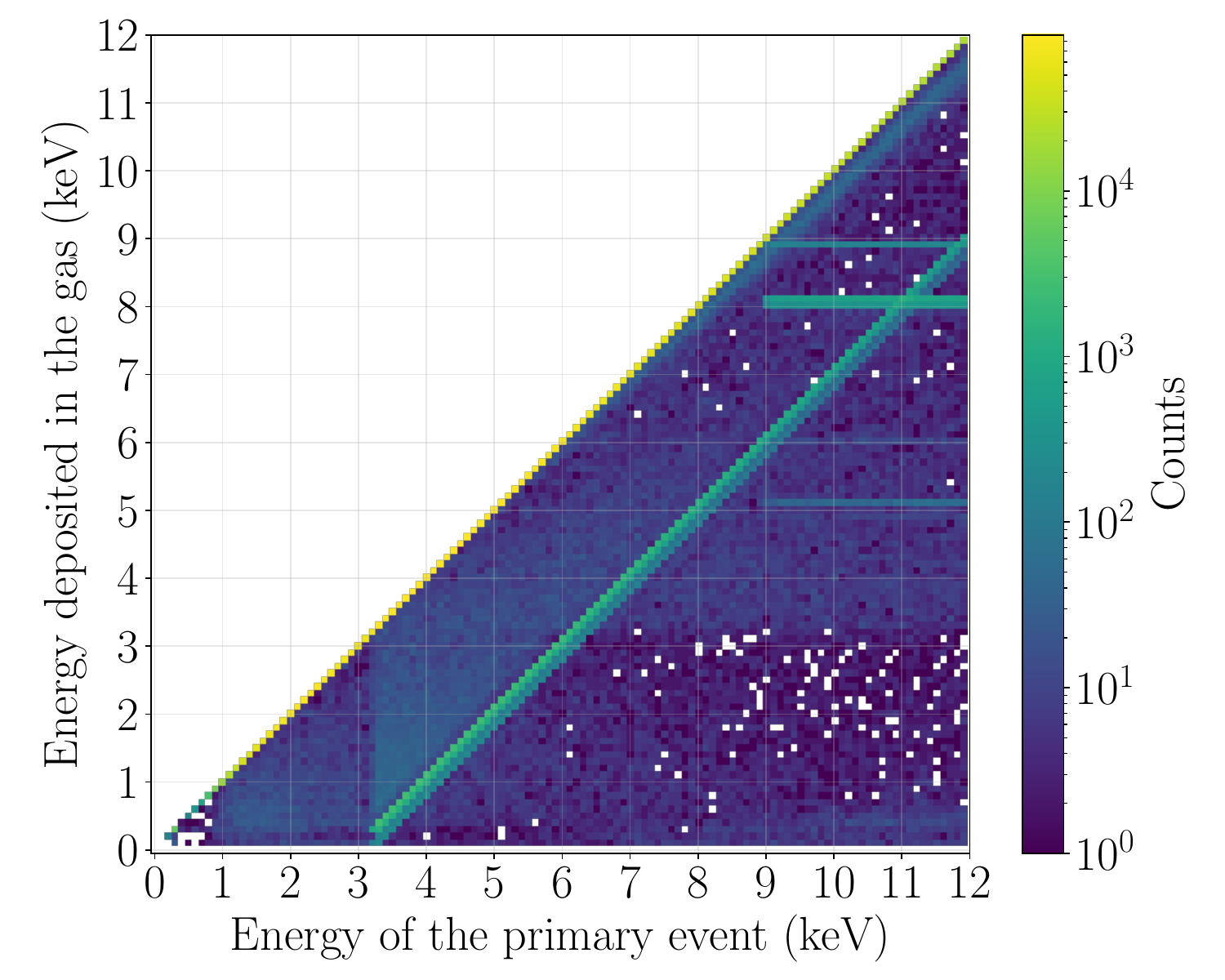}%
\caption{\label{fig:ArDetRespCOLZ} Visualization of the detector response matrix with $0.1\times 0.1$\,keV$^2$ bins.}
\end{figure}

To consider also these events towards computing the efficiency, the output of the simulations is used to build a detector response matrix $\mathcal{M}_{DR}$ with 2-dimensional bins of size $0.1 \times 0.1$\,keV$^2$ shown in Fig. \ref{fig:ArDetRespCOLZ}, where the energy of primary events is mapped to the deposited energy. It also encodes the Mylar window transmission, the gas absorption and the already mentioned higher order energy shifting effects.
One can see that most of the events lie on the main diagonal, but there are also clear fluorescence lines from copper $\mathrm{K\alpha}$ and $\mathrm{K\beta}$ at $~8$ and $~8.9$\,keV that appear as horizontal lines, and an accumulation of events at any point $~3.2$\,keV below the main diagonal, being 3.2\,keV the binding energy of the innermost Ar electron. Once this matrix is built, it can be convoluted with the incoming flux, in our case given by the Primakoff spectrum folded with the axion-photon conversion probability $P_{a\rightarrow \gamma}$ and the optics efficiency $\epsilon_o$. This flux is thus a vector of the number of X-rays at energy $E_i$ reaching the detector. A linear combination of this vector and $\mathcal{M}_{DR}$ gives the detected number of events with their corresponding energies in the gas. 
Finally one needs to incorporate the software efficiency. Based on measurements taken at 6 different energies in the CAST X-ray lab, we were able to keep this efficiency between 80\% and 90\% in the energy range of interest, and between 60\% and 70\% in the lowest energy range (Fig. \ref{fig:Efficiencies}).

Regarding the detector efficiency, the use of Xe-based gas mixtures, which have an inherently higher efficiency than Ar, as well as the removal of a differential cold window which was not required for vacuum operation, have also helped to increase the overall setup efficiency. All these improvements have affected the final result in a positive way, pushing the upper limit on $g_{a\gamma}$ lower than it would have been possible with past strategies.

\begin{figure}
\includegraphics[width=0.5\columnwidth]{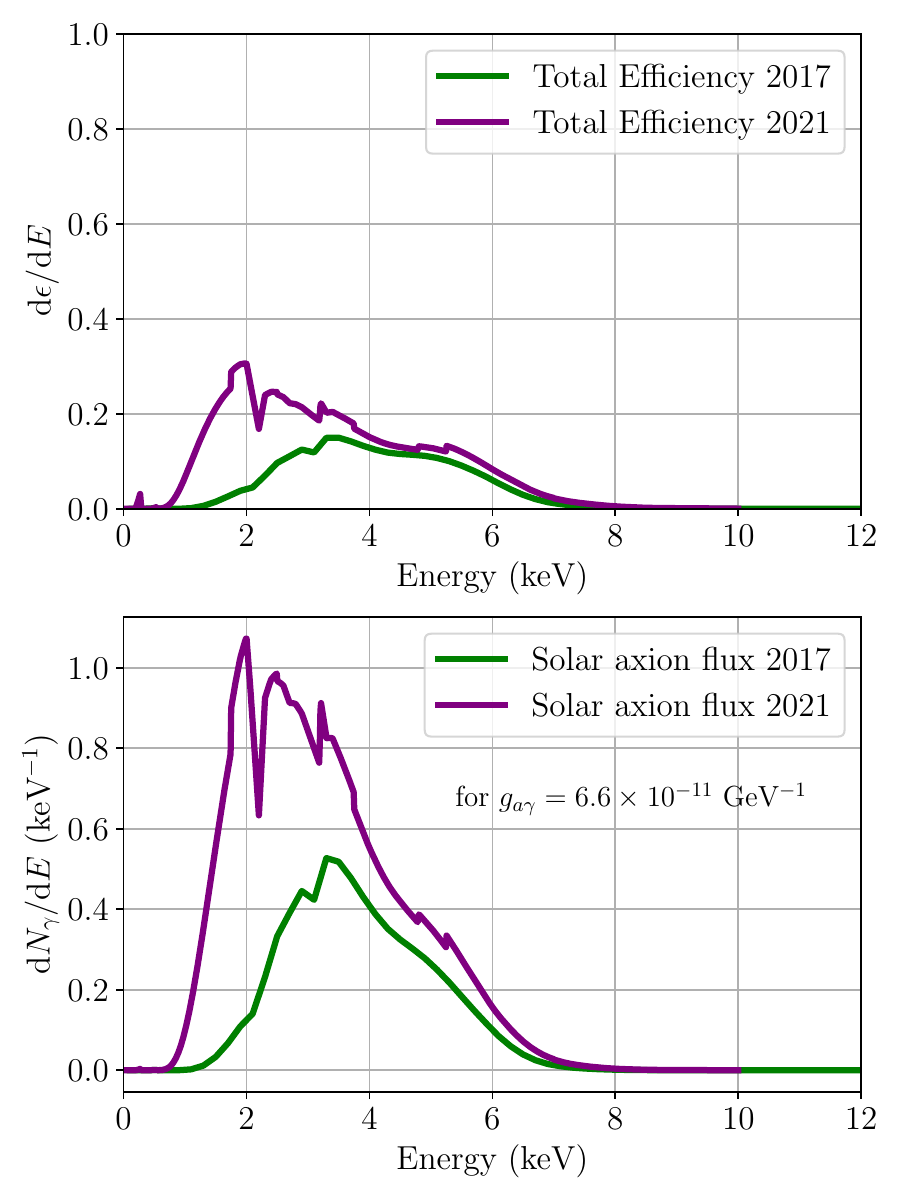}%
\caption{\label{fig:totalEfficiency} (top) Plot of the average total efficiency of the three datasets, where a horizontal line at d$\epsilon/$d$E=1$ represents 100\% efficiency (similar to Fig. \ref{fig:Efficiencies}). This is compared to the efficiency obtained in the 2017 data \cite{Anastassopoulos:2017ftlsupp}, highlighting our improved efficiency, especially at low energies where the solar axion flux is expected to peak. (bottom) Spectral distribution of the expected solar axion signal counts $N_\gamma$, corrected for efficiencies and integrated over time and area of the bore. The expected signal is twice as high due to the increased efficiency shown in the top panel.}
\end{figure}

The final efficiency is about twice as high as the efficiency obtained in the last results \cite{Anastassopoulos:2017ftlsupp}, and as seen in Fig.\ref{fig:totalEfficiency},  the improvement is especially noticeable at low energies where the expected solar axion flux is much higher. This has a significant impact in the final result.

\section{Systematic effects}
We estimate the systematic effects for the most relevant sources of uncertainty which we describe here. The results are listed in Table 
\ref{tab:systematics}.


\begin{table}[!b]
\renewcommand{\arraystretch}{1.2}
\setlength{\tabcolsep}{8pt}  
\begin{center}  
\begin{minipage}{0.6\textwidth}  
\caption{\label{tab:systematics} Systematic errors for different sources of uncertainty, quantified as the relative shift (upwards and downwards) of the upper limit on $g_{a\gamma}$ due to each particular source.}
\begin{ruledtabular}
\begin{tabular}{lcc}
Source of uncertainty & \multicolumn{2}{c}{Systematic effect in $g_{a\gamma}$} \\ \cline{2-3}
\hline
Magnetic field strength $B$ & $-0.16\%$ & $+0.22\%$ \\ 
Magnetic field length $L$ & $-0.27\%$ & $+0.27\%$ \\ 
Background level & $-0.22\%$ & $+0.22\%$ \\
Background area & $-1.56\%$ & $+0.22\%$ \\
Software efficiency & $-1.11\%$ & $+1.17\%$ \\
Spot position & $-0.22\%$ & $+0.12\%$ \\ 
Pointing accuracy & $-0.00\%$ & $+0.27\%$ \\ 
Theoretical axion flux & $-0.39\%$ & $+0.33\%$ \\
Solar model type & $-0.00\%$ & $+1.29\%$ \\ 
\hline
Total systematic uncertainty & $-2.00\%$ & $+1.85\%$ \\
\end{tabular}
\end{ruledtabular}
\end{minipage}
\end{center}
\end{table}

\textit{Magnetic field $B$ and magnet length $L$ --}
The CAST magnet was run at a very stable current of 13\,kA with negligible deviations from that value. The corresponding magnetic field is obtained by fitting data relating current and magnetic field (see e.g. \cite{JVogelTesissupp}). A linear fit is performed and the error in the fit parameters is used to compute the uncertainty by error propagation taking correlation into account. The estimated uncertainty is $B=8.805\pm 0.034$\,T.
The nominal value for the magnet length provided by the magnet group is $L=(9.26 \pm 0.05)$\,m. A 0.93346\,m long haloscope cavity was installed inside the bore, reducing the effective magnetic field length.

\begin{figure}
\includegraphics[width=0.5\columnwidth]{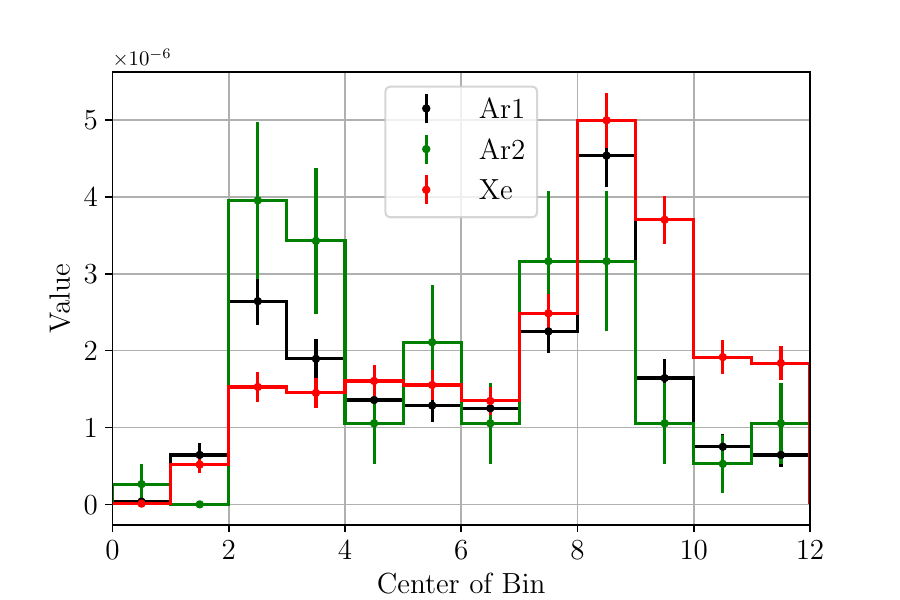}%
\caption{\label{fig:backgroundSystematics}Background level for the central region with $r=10$\,mm in keV$^{-1}$\,cm$^{-2}$\,s$^{-1}$ for Ar dataset 1 in black, Ar dataset 2 in green and and Xe in red. The error bars were used to compute the possible deviations.}
\end{figure}

\textit{Background level --}
The background level used for the calculations corresponds to the central readout area circle with $r=10$\,mm, with a bin width of 1\,keV and errors $\sqrt{N}$, where $N$ is the number of background counts. The background spectra that were included in the limit calculation are shown in Fig. \ref{fig:backgroundSystematics}. The systematic uncertainty is computed by considering the lowest/highest possible background, which is estimated by redefining the bin heights to match the lower/upper error bar endpoints.
We assume that the background is spatially uniform, but if we consider only a small area this assumption might not hold due to the low counts. Therefore, the background definition will be affected by the considered area. We thus compute the uncertainty in the result caused by using the background level of the inner readout area circle with $r=4$\,mm, and applying it to that same area and to the nominal $r=10$\,mm area.

\textit{Software efficiency --}
An algorithm has been designed to define the X-ray cuts for different energy ranges, which also affects the software efficiency in a positive way. This method yields a higher efficiency than other approaches used in the past and it maximizes the figure of merit. The results have been tested using the 3\,keV Ar escape peak and were found to be consistent, so a conservative 5\% uncertainty in the software efficiency is taken into account.

\begin{figure}
\includegraphics[width=0.5\columnwidth]{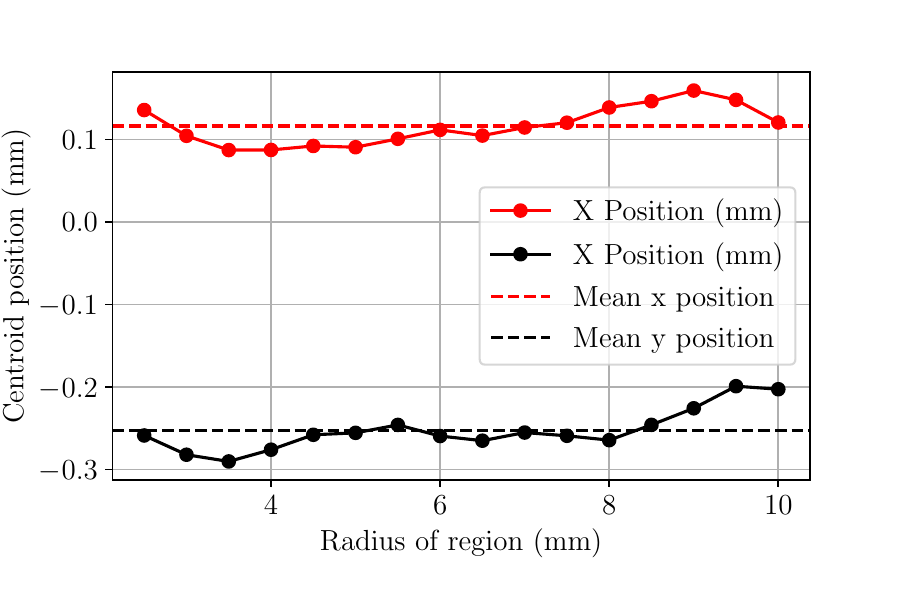}%
\caption{\label{fig:Centroid}Position of the centroid for different considered areas. The mean value was considered nominal, and the deviations were used to compute the uncertainty.}
\end{figure}

\textit{Alignment --}
The center of the spot was defined by computing the centroid of the data taken with the X-ray calibration source at CAST. The variation of the position of the centroid for circular readout areas with radii in the range $r=(2.5,10$)\,mm is shown in Fig. \ref{fig:Centroid}. The standard deviation of the centroid is 0.025\,mm, which produces a negligible change in the limit. Therefore, a more conservative value of $0.1$\,mm, based on the range of the centroid values, was considered.

\textit{Pointing accuracy--}
The pointing accuracy of the CAST experiment is 0.01$^{\circ}$. The CAST bore has an angular size of 0.5$^{\circ}$ while the solar axion signal comes from the inner 20\% of the Sun, resulting in 0.1$^{\circ}$. Thus, the solar core is always contained within the bore aperture. According to measurements taken in PANTER on- and off-axis, this results in a negligible displacement of the centroid on the readout plane. However, displacement can lead to a loss of up to 1\% efficiency.

\textit{Solar axion flux--}
Uncertainties in the theoretical solar axion flux and their effects in solar axion searches are studied in \cite{2101.08789supp}. The flux can be expected to have a statistical fluctuation of $\sim 1.5\%$, and it is also model dependent. Helioseismological models, based on the internal structure and dynamics of the Sun using information from internal sound waves, yield Primakoff fluxes that are consistently $\sim 5\%$ higher than those predicted by photospheric models. As axions are expected to be produced in the inner 20\% of the Sun, the former type of model is more appropriate, and thus we compute the model uncertainty by considering a flux 5\% lower. The particular expression used in this work is derived in \cite{Andriamonje:2007ewsupp}.

\textit{Total uncertainty--}
As the uncertainty sources are independent, the total uncertainty is calculated simply as the root sum of squares \smash{$\sqrt{\sum_{i=1}^{n} u_i^2}$} of the uncertainties $u_i$ of the $i$ sources, and it is defined by the asymmetric range -2.00\% and +1.85\%.

The background definition, the solar axion flux and the software efficiency are the most relevant uncertainties, and the rest are mostly negligible.
Still, all these contributions are minor compared to the statistical error inherent to rare-event experiments.

\section{Expected sensitivity}
\begin{figure}[!ht]
\includegraphics[width=0.5\columnwidth]{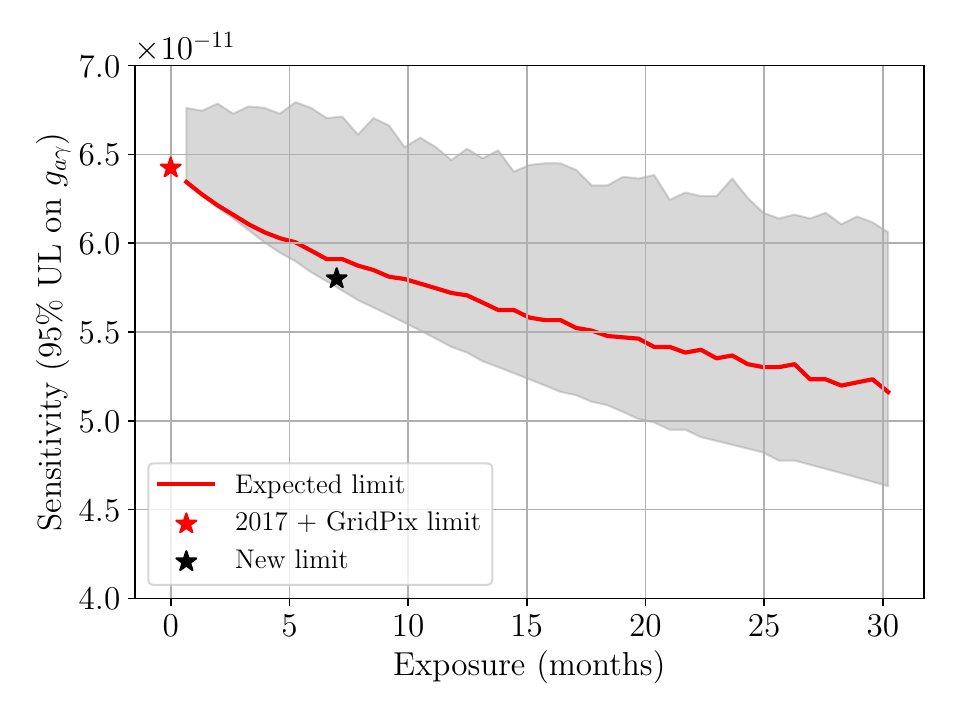}
\caption{\label{fig:sensitivity}Expected sensitivity of the experiment using Monte Carlo simulations given the current setup, background level, efficiencies, etc., and starting from the limit already set by the 2013-2018 data taking campaigns. The red line represents the expected limit computed as the median of the 1000 limits simulated in steps of 10 days of data taking. The time represented in the x-axis is the data acquisition time at CAST and the sensitivity is computed assuming 1.5 h of axion sensitive conditions per day. The grey shaded area is the band of possible limits for each exposure time, where the upper bound is the $99^{\mathrm{th}}$ percentile. }
\end{figure}

To compute the expected sensitivity of the experiment one needs to simulate sets of candidates (i.e., X-ray like events during tracking) that follow the background distribution in a given readout area, for a given exposure time and energy interval. In a binned case, a representative dataset such as the Asimov dataset \cite{Cowan_2011supp} provides an efficient solution to compute an expected limit. However, in an unbinned case a set of simulations has been performed instead, which also allows to study the possible variation due to statistical fluctuations. The expected sensitivity of the experiment can be simulated as follows:
\begin{itemize}
    \item For each exposure time step, draw a set of candidates that follow a Poisson distribution based on the background rate. This step is done 1000 times in time steps of 10 days, assuming 1.5\,h of data taking under axion sensitive conditions per day.
    \item Compute the limit that each of these set of candidates would produce.
    \item For each exposure time step, compute the median of the simulated limits as the expected limit.
\end{itemize}
The results of the simulations are shown in Fig. \ref{fig:sensitivity}, where the red line is the expected limit for a given exposure time and the grey band represents the range of possible values. In the current case, the expected sensitivity for 7 months of data taking is $g_{a\gamma}<0.59\times10^{-10}$\,GeV$^{-1}$, which is in agreement with our result.

\end{document}